\documentclass[fleqn,usenatbib]{aa}

\usepackage{newtxtext,newtxmath}
\usepackage[T1]{fontenc}
\usepackage{ae,aecompl}

\usepackage{graphicx}	
\usepackage{amsmath}	
\usepackage{amssymb}	
\usepackage{breakurl}
\usepackage{gensymb}    
\usepackage{threeparttable}
\usepackage{multirow}
\usepackage{color}

\newcommand{\prefactor}{\textsc{prefactor}\,}
\newcommand{\dysco}{\textsc{dysco}\,}
\newcommand{\pybdsf}{\textsc{pybdsf}\,}
\newcommand{\wsclean}{\textsc{wsclean}\,}
\newcommand{\clean}{\textsc{clean}\,}

\begin{document}
   \title{The resolved jet of 3C~273 at 150 MHz}
   \subtitle{Sub-arcsecond imaging with the LOFAR international baselines}

   \author{J.~J. Harwood\inst{1}\thanks{jeremy.harwood@physics.org}, S. Mooney\inst{2}\thanks{sean.mooney@ucdconnect.ie}, L.~K. Morabito\inst{3,4}, J. Quinn\inst{2}, F. Sweijen\inst{5}, C. Groeneveld\inst{5},\\E. Bonnassieux\inst{6,7}, A. Kappes\inst{8} \and J. Moldon\inst{9}
          }

    \institute{Centre for Astrophysics Research, School of Physics, Astronomy and Mathematics, University of Hertfordshire, College Lane, Hatfield, Hertfordshire AL10 9AB, UK
              \and
              School of Physics, University College Dublin, Belfield, Dublin 4, Republic of Ireland
              \and
              Centre for Extragalactic Astronomy, Department of Physics, Durham University, Durham DH1 3LE, UK
              \and Institute for Computational Cosmology, Department of Physics, University of Durham, South Road, Durham DH1 3LE, UK
              \and
              Leiden Observatory, Leiden University, P.O. Box 9513, 2300 RA Leiden, The Netherlands
              \and
              Universita di Bologna, Via Zamboni, 33, 40126 Bologna BO, Italy
              \and
              INAF, Instituto di Radioastronomia, Via Piero Gobetti, 101, 40129 Bologna BO, Italy
              \and
              Institut f\"ur Theoretische Physik und Astrophysik, Universit\"at W\"urzburg, Emil-Fischer-Str. 31, 97074 W\"urzburg, Germany	
              \and
              Instituto de Astrof\'isica de Andaluc\'ia (IAA, CSIC), Glorieta de las Astronom\'ia, s/n, E-18008 Granada, Spain
             }

\date{Received XXX, 2021; accepted XXX}

\label{firstpage}
\abstract
{
Since its discovery in 1963, 3C~273 has become one of the most widely studied quasars with investigations spanning the electromagnetic spectrum. While much has therefore been discovered about this historically notable source, its low-frequency emission is far less well understood. Observations in the MHz regime have traditionally lacked the resolution required to explore small-scale structures, such as knots and diffuse jet emission, that are key to understanding the processes that result in the observed emission. Advances in the processing of LOFAR international baseline data have now removed this limitation, providing the opportunity to explore this key area for the first time.
}
{
In this paper we use the first sub-arcsecond images of 3C~273 at MHz frequencies to investigate the morphology of the compact jet structures and the processes that result in the observed spectrum. We will determine the jet's kinetic power, place constraints on the bulk speed and inclination angle of the jets, and look for evidence of the elusive counterjet at $150$ MHz. 
}
{
Using the full complement of LOFAR's international stations (German, Poland, France, UK, Sweden), we produce $0.31 \times 0.21$ arcsec images of 3C~273 at $150$ MHz. Using ancillary data at GHz frequencies, we fit free-free absorption (FFA) and synchrotron self-absorption (SSA) models to each region in order to determine their validity in explaining the observed spectra.
}
{
The images presented display for the first time that robust, high-fidelity imaging of low-declination complex sources is now possible with the LOFAR international baselines. We show that the main small-scale structures of 3C~273 match those seen at higher frequencies, with a tenuous detection of an extension to the outer lobe. We find that FFA and SSA models are able to describe the spectrum of the knots and, while differentiating between model types requires further observations, conclude that absorption is present in the observed emission. We determine the kinetic power of the jet to be in the range of $3.5 \times 10^{43}$ - $1.5 \times 10^{44}$ erg s$^{-1}$ which agrees with estimates made using higher frequency observations. We derive lower limits for the bulk speed and Lorentz factor of $\beta \gtrsim 0.55$ and $\Gamma \geq 1.2$ respectively. The counter-jet remains undetected at $150$ MHz, placing a limit on the peak brightness of $S_\mathrm{cj\_150} < 40$ mJy beam$^{-1}$.
}
{}


\authorrunning{Harwood et al.}
\titlerunning{The resolved jet of 3C~273 at low frequencies}
\maketitle

\graphicspath{{./images/}}

\section{Introduction}
\label{intro}

As the first quasars discovered, 3C~273 (J1229+0203) is one of the most well-studied objects of its type at almost every waveband with over a thousand peer-reviewed publications aiming to understand its properties. At its centre, 3C~273 is observed to host a supermassive black hole (SMBH) of mass $(890 \pm 190) \times 10^{6}$ M$_{\odot}$ \citep{2004ApJ...613..682P} with a complex one-sided jet observed on parsec scales \citep{2020AdSpR..65..725Z}. The possibility that 3C~273 is intrinsically one-sided was considered during early investigations \citep{1985Natur.318..343D} but, given AGN unification theory \citep{1995PASP..107..803U} and observations of other AGN, there is a general consensus in recent literature that 3C~273 is two-sided and intrinsically symmetrical observed at an angle of $\approx 5 \degree$ \citep{2016ApJ...818..195M} with a jet-to-counterjet flux ratio greater than $5300$ \citep{2001ApJ...549L.161S}.

\begin{figure*}
    \centering
    \includegraphics[width=\columnwidth]{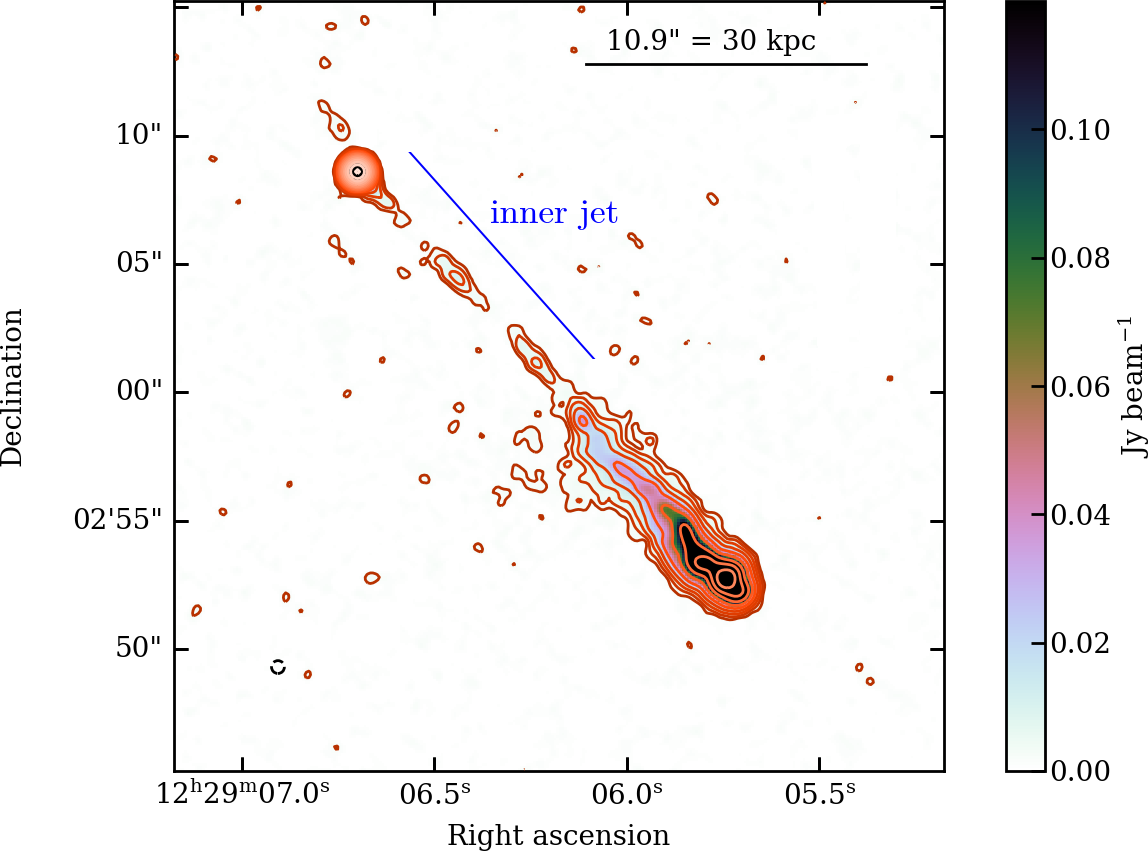}
    \includegraphics[width=\columnwidth]{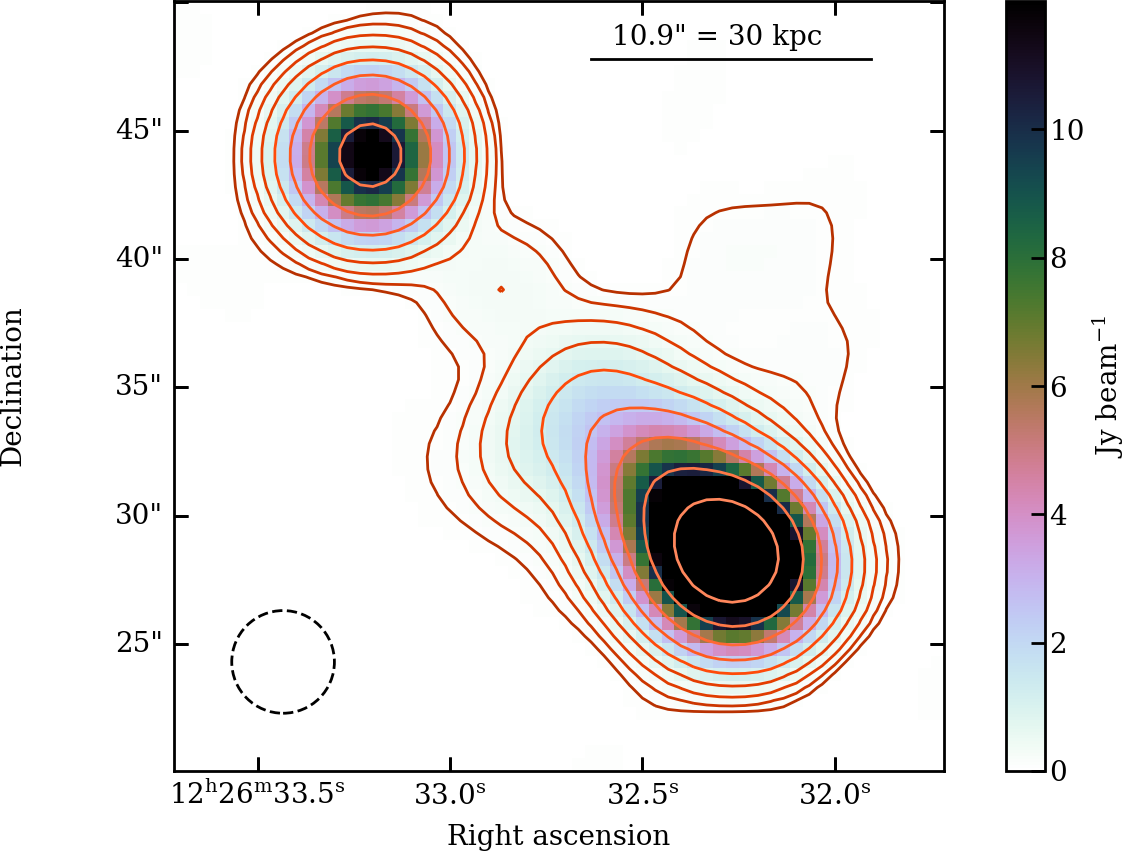}
    \caption{Left: VLA C-band of 3C~273 at $4885$ MHz at a resolution is $0.5$ arcsec. Contours are such that $S_{cont} = 2\times2^{n}$ mJy beam$^{-1}$ with a peak brightness of $35.99$ Jy beam$^{-1}$. The inner jet is marked for clarity. Right: VLA P-band of 3C~273 at $327$ MHz at a resolution of $0.5$ arcsec. Contours such that $S_{cont} = 80\times2^{n}$ mJy beam$^{-1}$ with a peak brightness of $36.37$ Jy beam$^{-1}$. These data are reproduced from \citet{2017A&A...601A..35P}.}
    \label{fig:3C273-two-vla}
\end{figure*}

The core flux density is known to be highly variable at and above GHz frequencies with the flux density at $37$ GHz varying in the range of $10 - 57$ Jy \citep{2016Ap.....59..213S}. Intraday variability has also been detected \citep{2015MNRAS.451.1356K}, but no studies to date have been able to monitor time variability in the megahertz regime due to the limitations currently imposed at these frequencies, e.g. resolution, artefact suppression, flux scale accuracy. The radio emission from the core is understood to be optically-thick synchrotron radiation and, despite the observed variability, the radio spectrum is known to be flat between $0.1$ and $100$ GHz.

On larger scales, a faint inner jet extending $12$ arcsec ($33$ kpc) joins the nucleus to an outer jet \citep{1985Natur.318..343D,1993A&A...267..347C} which can be seen in the Karl G. Jansky Very Large Array (VLA) observations shown in Fig.~\ref{fig:3C273-two-vla}. In addition to the radio waveband, this inner jet has been detected at optical wavelengths by the Hubble Space Telescope (HST; \citealp{2003AJ....125.2964M}). In contrast, the outer jet is bright and highly polarised, extending $27$ kpc in the plane of the sky \citep{2017A&A...601A..35P}. The jet has a knotty structure at radio, optical, and X-ray frequencies indicating modulation in the jet's kinetic power. \citet{2004ApJ...604L..81G} also note a decrease in the ratio of the X-ray to radio brightness with distance from the core, that is, as the radio jet increases in brightness with distance from the core, the X-ray jet becomes fainter.
 
Despite extensive research, there remains no consensus regarding the origin of the X-ray emission associated with the relativistic jet \citep{2006ApJ...653L...5G} and there are varying interpretations of the multiwavelength spectra for the knots in the jet of 3C~273. However, it is generally agreed that for each knot, a population of relativistic electrons gives rise to the radio-to-infrared emission via the synchrotron process. Radio and optical polarisation measurements are consistent in magnitude and orientation, which is further evidence that the emission across these wavebands is non-thermal in origin and stems from a single population of electrons \citep{1991A&A...252..458R}. However, the mechanism behind the ultraviolet-to-X-ray emission, and how it fits into the overarching multi-wavelength picture, is less well understood.

While X-ray emission from Fanaroff and Riley class I radio galaxies (FR-I; \citealp{1974MNRAS.167P..31F}) and BL Lacs is usually explained by extrapolating the radio-to-infrared synchrotron spectrum, in general the X-ray flux of FR-IIs and quasars are higher than expected. Observations of the knots in the jet show that 3C~273, a quasar believed to have an underlying FR-II morphology, is no exception with simple power law models underestimating the X-ray flux. To account for this shortfall, the favoured model for the X-ray emission has typically been inverse-Compton scattering of the cosmic microwave background (IC/CMB; \citealp{2001ApJ...549L.161S}). However, several challenges to this interpretation have been made. Firstly, models seeking to reproduce the X-ray flux solely via IC/CMB require relativistic bulk motion on kiloparsec scales. Superluminal speeds (${\sim}10$c) on parsec scales near the core as measured by Very Long Baseline Interferometry (VLBI) studies open the door to this possibility, but whether these speeds are maintained on larger scales is unclear. \citet{2004MNRAS.351..727A} suggest that jets should decelerate to mildly relativistic speeds on kiloparsec scales due to the entrainment of ambient gas, a finding later supported by the modelling of \citet{2009MNRAS.398.1989M} who verify that jets decelerate, constraining the bulk Lorentz factor, $\Gamma$, for kiloparsec-scale radio jets of quasars to $1.18 < \Gamma < 1.49$.

Ultimately, an IC/CMB origin of the X-ray emission was effectively ruled out by observational $\gamma$-ray results. \citet{2015ApJ...805..154M} showed that IC/CMB models violate the upper limits on the $\gamma$-ray flux of 3C~273 and PKS 0637-752 at the $99.99$ per cent confidence level in more than one Fermi-LAT energy band. These deep upper limits at GeV energies constrain the Doppler beaming factors, implying that the jet is not highly relativistic on kiloparsec scales as is required for IC/CMB models.

\begin{figure*}
    \centering
    \includegraphics[width=\textwidth]{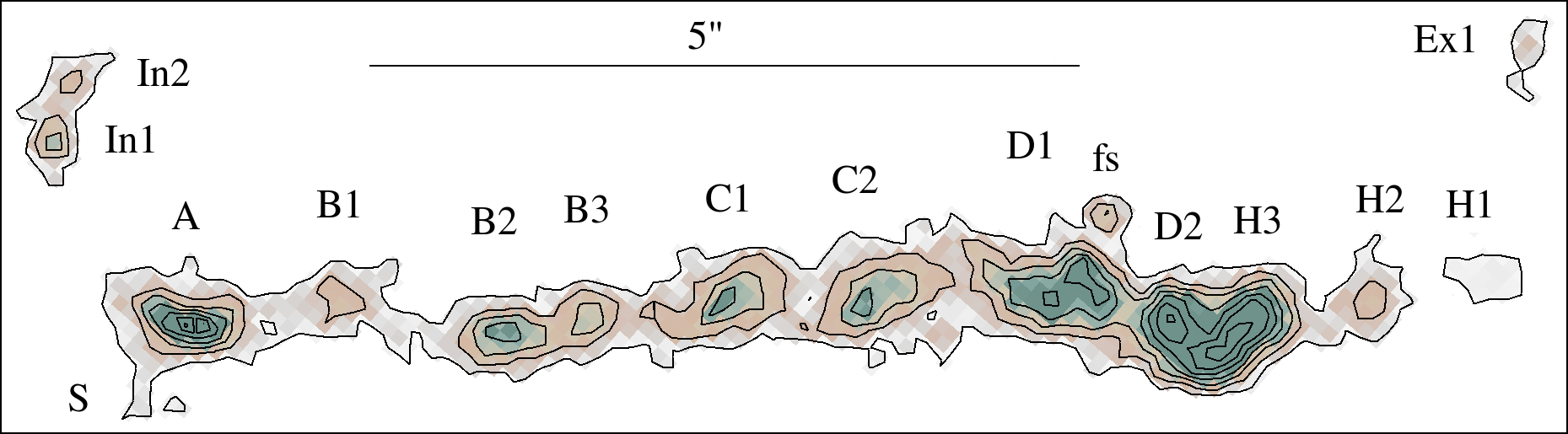}
    \caption[Outer jet of 3C~273]{HST image of the 3C~273 outer jet masked at $0.7$ e$^{-}/\,$s to only show the jet. The direction of motion is left-to-right, with the core ${\sim} 11$ arcsec to the left of the image. In1, In2, and Ex1 are galaxies unrelated to the jet; fs is a foreground star. The naming convention is from \citet{2005A&A...431..477J}, with the addition of features fs, Ex1, and H1.}
    \label{fig:3C273-knot-names-border}
\end{figure*}

One alternative to the upscattering of the CMB is the synchrotron self-Compton (SSC) processes that may also plausibly produce X-ray emission in the jet. However, \citet{2001ApJ...549L.161S} find that for the X-ray component to be more luminous than the radio component, as is observed, weak magnetic fields ($B \lesssim 10^{-3}$ mG) orders of magnitude below the equipartition value are required, unless the jet is significantly debeamed ($\delta \approx 0.3$--$0.5$). Such significant debeaming is incongruous with the jet-to-counterjet flux ratio of $5300$ \citep{2001ApJ...549L.161S} and so SSC is unlikely to be solely sufficient to produce the X-ray fluxes that are observed. A recent study by \citet{2020ApJ...893...41W} comprehensively demonstrated that the synchrotron, IC/CMB, and SSC components of a single population of electrons are unable to reproduce the expected X-ray emission. While IC/CMB and SSC are therefore likely occur at some level in the jet of 3C~273, they are not the primary mechanisms behind the X-ray emission. Given that such `one-zone' models struggle to explain high-energy emission not only in knots, but in the jets of FR-Is more generally (e.g. \citealp{2012MNRAS.423.1368H} and references therein), more complex models are likely required if one is to accurately describe the processes that drive the observed emission.

Many competing hypotheses have been proposed to resolve these inconsistencies. \citet{2013ApJ...773..186C} favour a second population of particles, whether leptonic or hadronic, to explain the X-ray emission for the knots. \citet{2020ApJ...893...41W} also found that the multiwavelength spectra can be accurately reproduced by including synchrotron processes associated with a second population of particles. It is unclear, however, whether this second population consists of electrons or protons, as the SED predictions in both cases satisfy the observations they considered. \citet{2002MNRAS.332..215A} proposed a broken power law spectrum of accelerated protons from e.g. the jet shear boundary, producing synchrotron radiation as the origin of the X-ray and $\gamma$-ray flux from the knots. However, super-Eddington jet powers or strong (${\sim}10$ mG) magnetic fields are required in the hadronic case as more energy is needed to accelerate the protons sufficiently. A secondary leptonic population would overcome this issue, although the origin of such a second population is unknown.

It has also been suggested that two populations of electrons could have different acceleration mechanisms. \citet{2015ApJ...806..188L} proposed separate shock regions in the knots for the radio and X-ray emission with distinct populations formed upstream and downstream because of the shock compression effect, where the maximum energy of the injection spectrum into the downstream emission region is lower than the maximum energy of the upstream region. Assuming that particle acceleration in the knots occurs in situ (i.e. that the acceleration and emission zones are co-spatial), testing whether the radio and X-ray knots are spatially coincident can provide evidence of multiple acceleration sites, and hence multiple populations. Indeed, \citet{2017ApJ...844...11M} found that the X-ray knots are upstream of the radio knots by $0.2 - 1.0$ kpc. Alternatively, \citet{2017ApJ...842...39L} have proposed shear acceleration to produce high energy electrons, but a rigorous comparison between predictions of such models has yet to be performed.

As highlighted by \citet{2001ApJ...549L.161S}, an underlying issue exists with the ad hoc nature of invoking a second component in that it produces a scenario where there are enough free parameters that the physical state of the jet is essentially unconstrained without comprehensive observations, as complex models invariably provide better fits to data due to over-fitting. Therefore despite extensive research, there remains no consensus regarding the origin of the X-ray emission associated with the relativistic jet.

One area of investigation that has yet to be fully explored is that of low-frequency emission. Historically, observations in the MHz regime have lacked either the resolving power and/or UV coverage to investigate the knot and diffuse jet emission at long wavelengths. However, the international baselines provided by the LOw-Frequency ARray (LOFAR; \citealp{2013A&A...556A...2V}) provide the opportunity to investigate this previously unexplored area for the first time, and provide a baseline reference from which to measure the proper motion of the knots at low frequencies. In this paper, we will present high-resolution images at $150$ MHz, the first sub-arcsecond images of 3C~273 below $1$ GHz, to investigate five key unresolved questions:

\begin{enumerate}
\item What compact structures are observable in the jet of 3C~273 at low-frequencies?\\
\item What processes result in the observed low-frequency emission of the jet knots?\\
\item How does the jet's kinetic power determined at low frequencies compare to previous estimates?\\
\item Is the previously unseen counterjet detected at low-frequencies?\\
\item What constraints can be placed on the bulk speed of the jets and their inclination angle?
\end{enumerate}
\vspace{-2mm}

\noindent In identifying the features of the jet, we adopt the naming convention of \citet{2005A&A...431..477J}, with the addition of features fs, Ex1, and H1 (Fig.~\ref{fig:3C273-knot-names-border}). Throughout this paper, we define the spectral index such that $S \propto \nu^{-\alpha}$ and use a concordance model in which $H_0=71$ km s$^{-1}$ Mpc$^{-1}$, $\Omega _m =0.27$ and $\Omega _\Lambda =0.73$ \citep{spergel03}.

\section{Observations and data reduction}
\label{method}

\subsection{LOFAR observations} \label{sec:observations}

3C~273 was observed during LOFAR cycle 8 using $24$ core stations, $14$ remote stations, and $12$ international stations (six stations in Germany, three in Poland, and one in each of France, the UK, and Sweden). The observations were made using a 4-hour track when 3C~273 was highest in the sky. This is particularly desirable when using LOFAR due to the source's low declination ($+02\degree$), where increased ionospheric decorrelation of the signal increases the difficulty in achieving a robust calibration of the source. Since 3C~273 is radio-bright, these shorter observations will provide sufficient signal-to-noise to image the source with the added benefit of increasing the likelihood of observing during optimal ionospheric conditions.

3C~295 and 3C~280 were chosen as calibrators. While 3C~295 and 3C~280 are $50\degree$ and $46\degree$ from 3C~273 respectively, as they were only used to calibrate the direction-independent effects it does not significantly impact the overall robustness of the calibration. 3C~295 is a bright standard flux density calibrator at low frequencies \citep{2012MNRAS.423L..30S} and situated at a high declination. It is therefore an ideal calibrator source. A summary of the observations, including the observations of the two calibrators, is given in Table~\ref{tab:lofar-observations}.

\subsection{Preprocessing} \label{sec:preprocessing}

The raw data were initially processed by the observatory in the standard manner. The preprocessing pipelines automatically flagged RFI, averaged the data in time and frequency, and performed demixing of the so-called A-team sources (Cassiopeia A, Cygnus A, Hydra A, Taurus, Virgo A) where required. Due to the large size of LOFAR observations ($4 - 20$ terabytes per pointing), the data were averaged from the raw time-resolution of one second and $64$ ch/sb (channels per subband) to 2 seconds and $16$ ch/sb providing a bandwidth of $12.2$ kHz per channel. This reduced the computational requirements required for the data reduction and analysis while having negligible impact on the science cases presented in this paper.

\begin{table}
    \centering
    \caption{Summary of the LOFAR observations}
    \label{tab:lofar-observations}
    \begin{tabular*}{\linewidth}{c@{\extracolsep{\fill}}cccc}
        \hline
        \hline
        Observation date&Duration&Source&Observation ID\\
        &(minutes)&& \\ 
        \hline
        13:13 23/08/2017 & $10$  & 3C~295 & L606008 \\
        13:24 23/08/2017 & $240$ & 3C~273 & L606014 \\
        17:25 23/08/2017 & $10$  & 3C~280 & L605068 \\
        \hline
    \end{tabular*}
    \vskip 5pt
    \begin{minipage}{\columnwidth}
        \small Summary of LOFAR observations at $134 - 164$ MHz under project ID LC8\_032. Observations were averaged to 2 seconds and 16 channels per sub-band during observatory preprocessing. Data are accessible at \url{https://lta.lofar.eu}
    \end{minipage}
\end{table}

\subsection{Calibrator processing for direction independent effects} \label{sec:core-and-remote-station-calibration}

Upon completion of the preprocessing, the data were downloaded and initial calibration performed using the LOFAR pre-facet calibration pipeline (\prefactor\footnote{\url{https://github.com/lofar-astron/prefactor}}; \citealp{2019A&A...622A...5D}) on the core and remote stations. As only one calibrator can by used by the pipeline, the \prefactor calibrator pipeline was first used to process the 3C~295 and 3C~280 data.

The solutions from the pipeline runs were then inspected. The LOFAR station bandpass solutions associated with the 3C~280 run had the incorrect shape compared to the theoretical bandpass (likely due to calibrator model inaccuracies), and so 3C~295 was used for the remainder of this analysis. A two-component model of 3C~295 is packaged with \prefactor which, while sufficient for studies that only use the core and remote stations, is not suitable for long-baseline analysis. A high-resolution model provided by the LOFAR VLBI working group, consisting of \clean components \citep{1974A&AS...15..417H} was therefore used.

\begin{figure*}
    \centering
    \includegraphics[width=\textwidth]{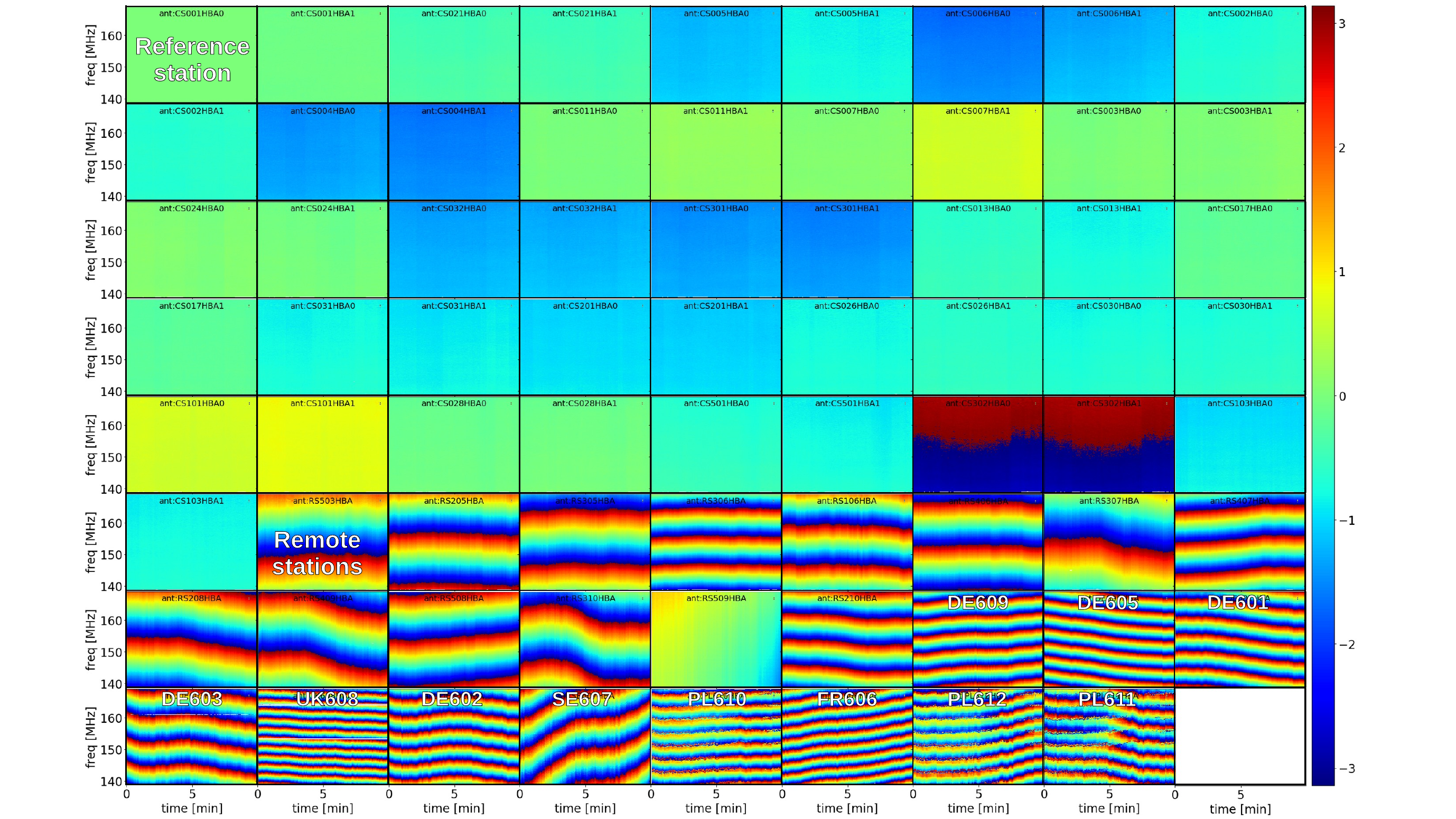}
    \caption{Calibrator phase solutions for 3C295 for the XX polarisation. The reference station is CS001HBA0 and the plots increase by baseline length from the reference station reading from left to right. All plots are on the same time and frequency axes. The first remote station is labelled, and individual international stations are labelled.}
    \label{fig:3C273-ph-polXX}
\end{figure*}

The observations were put through the \prefactor pipeline. The calibrator phase solutions are shown in Fig.~\ref{fig:3C273-ph-polXX} where CS001HBA0 is the reference station. All other core stations have small phase offsets with respect to this reference station. The remote stations show more variation than the core stations with the the international stations showing the fastest variations with frequency. The stations in this figure are sorted by baseline length from the reference station. An interactive map showing the station locations can be found online\footnote{https://www.astron.nl/lofartools/lofarmap.html} The data were then compressed using \dysco\footnote{\url{https://github.com/aroffringa/dysco}} \citep{2016A&A...595A..99O}, allowing LOFAR data to be compressed by approximately a factor of six with only a $1$ per cent increase in system noise for typical correlator time and frequency resolutions.

Before calibrating the international stations, the data were processed through the \prefactor target pipeline, which flags the data, removes contributions from bright off-axis sources, and calibrates the phases on the core and remote stations. Following this, the data are ready for international station calibration.

\begin{figure}
    \centering
    \includegraphics[width=\columnwidth]{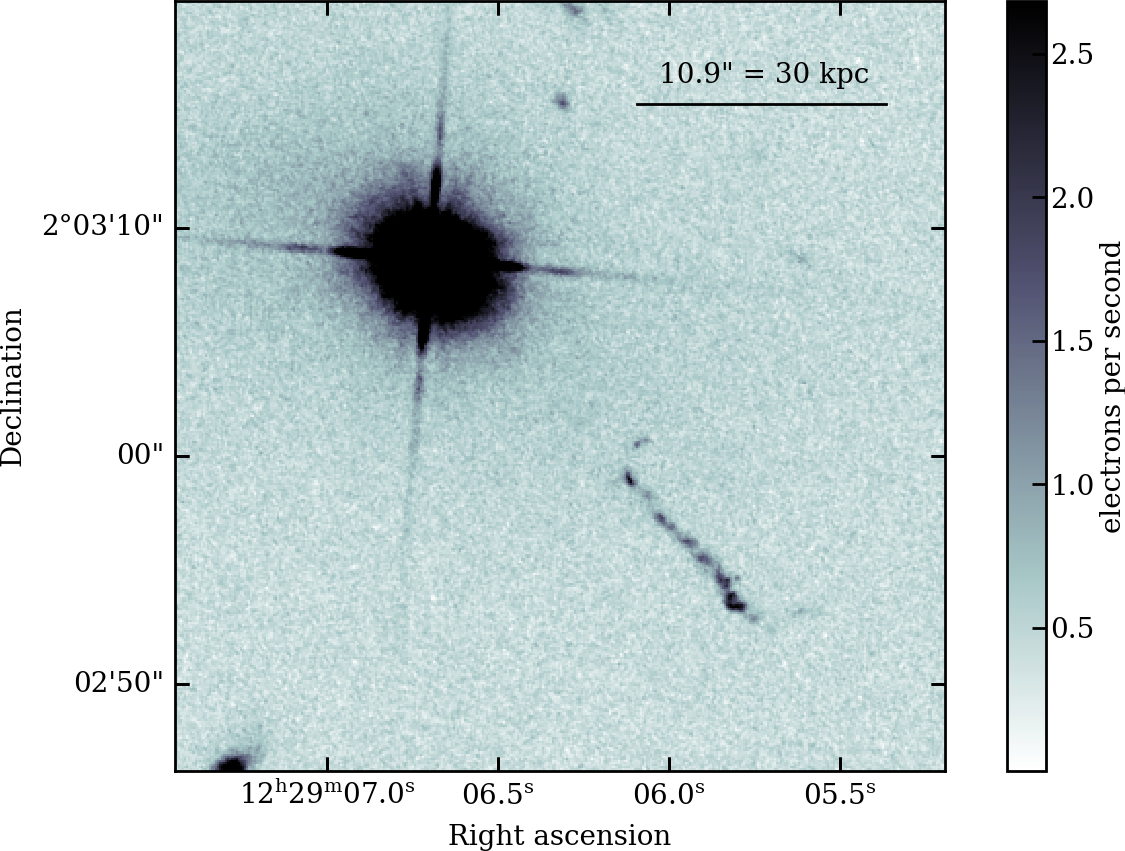}
    \caption{HST image of 3C~273 at $850 - 1700$ nm with a peak brightness of $28606$ e$^{-}$/s. The image combines four exposures providing a $37$ second integration time.}
    \label{fig:3C273-hst}
\end{figure}

\begin{figure*}
    \centering
    \includegraphics[width=\textwidth]{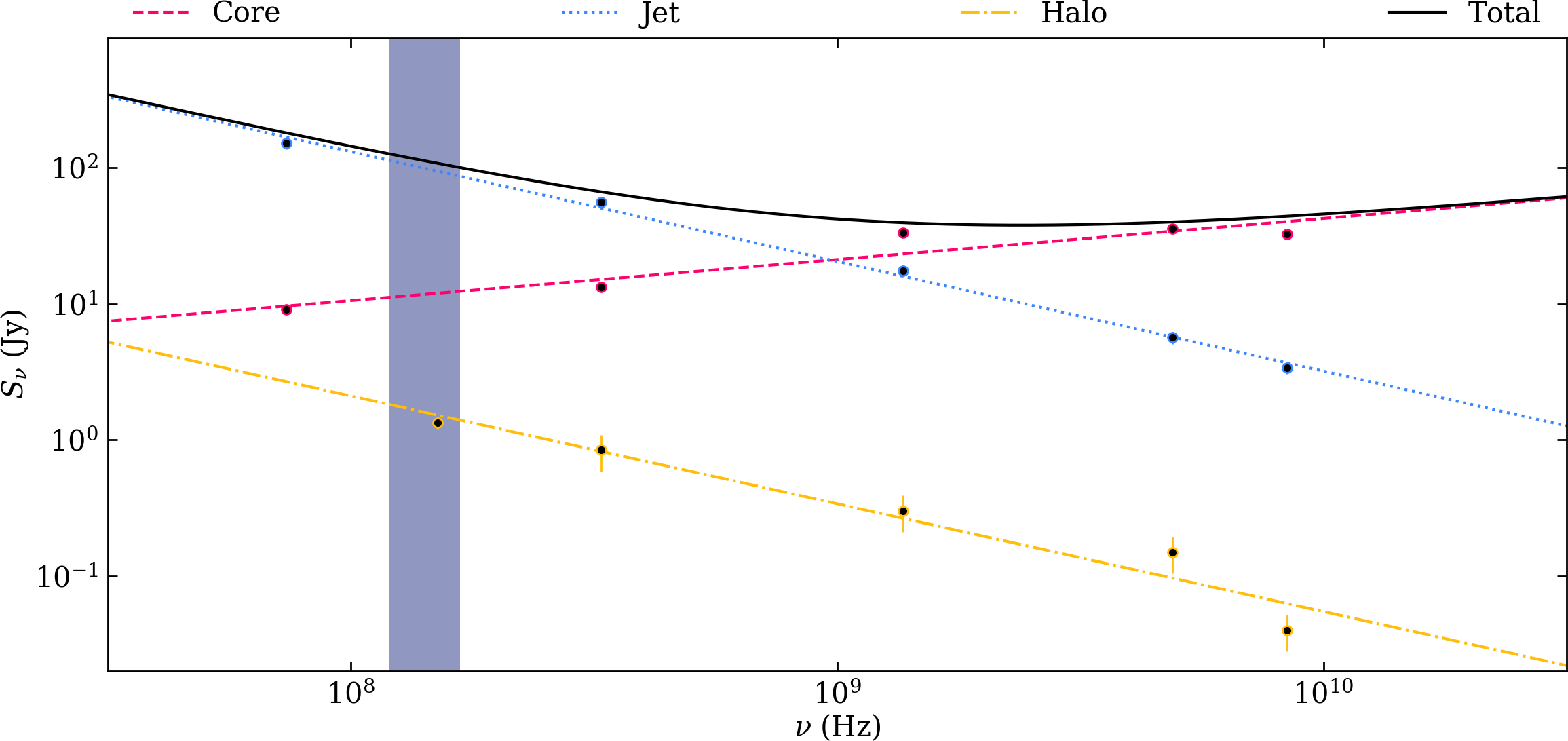}
    \caption{Components to the 3C~273 spectral energy distribution. Data are from \citet{2017A&A...601A..35P}, except for the $151$ MHz flux density of the halo, which is from \citet{2016ApJ...833...57P}. The flux density of the jet includes all knots. The halo refers to the diffuse cocoon of radio emission encompassing the jet (A, B, and R of Fig. \ref{fig:3C273-contours-on-vla}). The radio spectral indices of the core, jet, and halo are $0.30 \pm 0.07$, $-0.81 \pm 0.03$, and $-0.79 \pm 0.11$ respectively. The spectral indices of the jet and the halo are consistent with optically-thin synchrotron emission, as expected. The LOFAR HBA bandwidth is shown by the shaded region.}
    \label{fig:3C273-fluxes-2017AandA}
\end{figure*}

\subsection{International station calibration} \label{sec:international-station-calibration}

The LOFAR long-baseline pipeline \citep{2016A&A...595A..86J, morabito21, jackson21} was run to calibrate the international stations. The pipeline takes the 3C~273 data and the \prefactor solutions as input, specifically the gain solutions from the calibrator pipeline and the phase solutions from the target pipeline.

Based on a 6 arcsecond resolution image, the hotspot of 3C~273 (i.e. the jet terminus) is determined to have the largest integrated flux density. The pipeline automatically determined this to be the best in-field calibrator and after applying the \prefactor calibration solutions the data were phase-shifted to the position of this component. The core LOFAR station beams were then added coherently to form station ST001. A phase-only self-calibration of the in-field calibrator was then performed against an initial model. The pipeline, which is optimised to self-calibrate compact calibrators with simple structure from the LBCS (Jackson et al., 2021), attempts to initialise self-calibration with a point source. In this case, a single Gaussian model with major and minor full widths at half maxima (FWHM) of $0.1$ arcsec was initially used but the results were poor, given that a point source model neglects the large-scale jet. An initial model of the whole source was therefore built using \pybdsf\footnote{https://www.astron.nl/citt/pybdsf/} from a lower resolution image (${\sim}1$ arcsec) produced from four hours of LBA data \citep{groeneveld21}. All of the calibration solutions were then applied to the data.

\subsection{Self-calibration and imaging} \label{sec:self-calibration-and-imaging}

From the initial model for the brightness distribution of 3C~273 used in Section~\ref{sec:international-station-calibration} the relative gains (amplitude and phase) were calibrated. The calibration solutions were applied to the data and which was then imaged, providing an improved model for the observed sky surface brightness distribution. \wsclean was used for imaging stage \citep{2014MNRAS.444..606O} which uses both the multiscale clean algorithm (MSCLEAN; \citealp{cornwell08, 2011A&A...532A..71R}) to improve the diffuse emission fidelity, and multi-frequency synthesis (MFS; \citealp{2011A&A...532A..71R}) to account for spectral variations across the bandwidth. These steps were iterated in the standard manner until the image produced did not result in an improvement in quality\footnote{The absolute value of the peak brightness divided by the minimum brightness was used to measure image quality as both the peak brightness should increase and the minimum brightness should decrease as the calibration improves.}.

\subsection{Astrometry and flux-scale bootstrapping} \label{sec:astrometry-and-flux-scale-bootstrapping}

Attaining absolute astrometric positions with LOFAR is inherently difficult due to ionospheric refraction causing sources to shift by several arcseconds in the image plane at $150$ MHz. Accurate total electron content (TEC) solutions should mostly correct for this effect but, combined with other factors such as positional shifts during self-calibration, additional alignment was also performed to ensure robust astrometric positions were achieved.

In order to achieve robust astrometric positions, the core of 3C~273 was therefore aligned for the LOFAR, VLA, and HST observations to the position as reported by \citet{1995AJ....110..880J}, which is accurate to better than $3$ mas. For the radio data, the core position was taken to be the pixel of the peak brightness, and for the near-infrared data the intersection of the diffraction spikes was used to define the core position. For AGN, the position of the core can drift across frequencies; however, this effect is seen on milliarcsecond scales and so is unlikely to have a significant impact on our analysis. Given the resolution of the LOFAR, VLA, and HST data, we conclude the accuracy of the astrometry in the composite images is $<0.05$ arcsec.

Initially, the flux calibrator 3C~295 and the \citet{2012MNRAS.423L..30S} flux scale were used to set the flux scale in the LOFAR observations. However, imperfect beam models mean that the relative gains between 3C~295 and 3C~273 are not well determined, introducing an error on the absolute flux density scale. To determine the scaling factor required to correct for these errors, the flux measurements of \citet{2017A&A...601A..35P} and \citet{2016ApJ...833...57P} were used to perform a linear least squares regression and ascertain the spectral indices (Fig.~\ref{fig:3C273-fluxes-2017AandA}). \pybdsf was then run on the LOFAR image to model the source and to find the total flux density of the core component. The fit of the core at $150$ MHz was then used to bootstrap the flux scale in the LOFAR data.

\begin{figure*}
    \centering
    \includegraphics[width=\columnwidth]{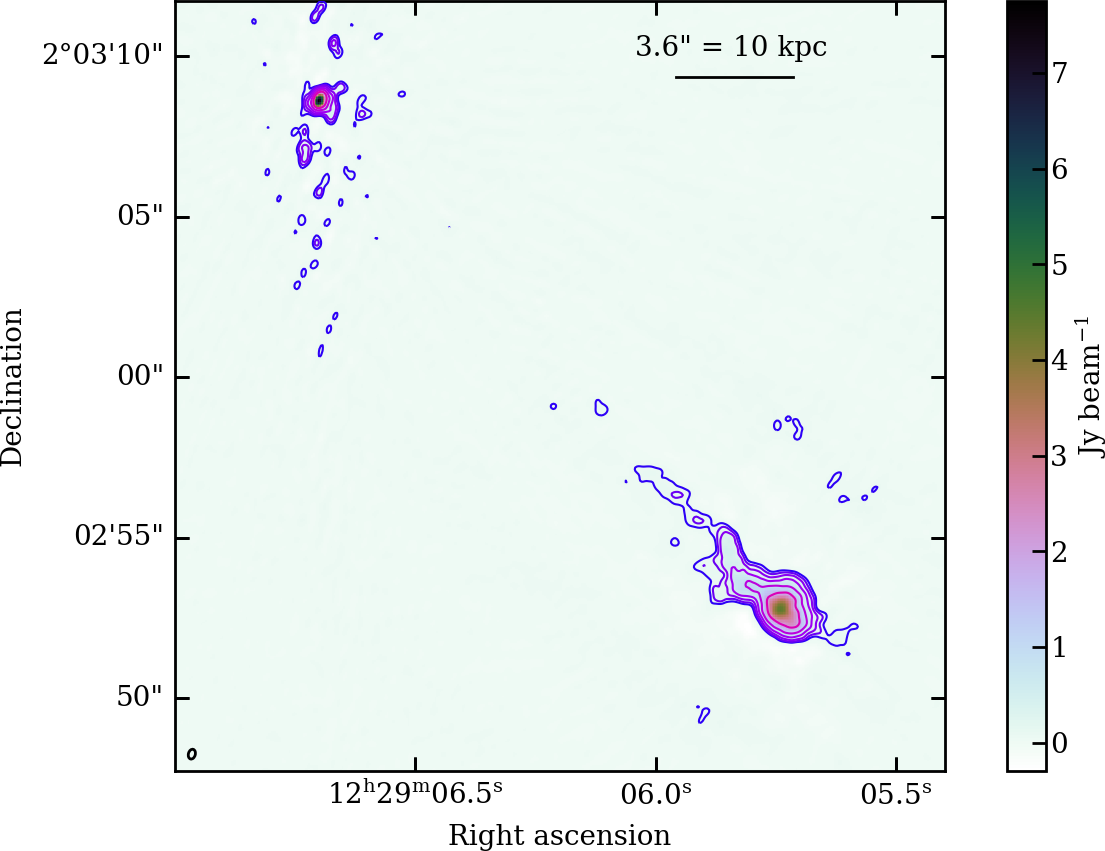}
    \includegraphics[width=\columnwidth]{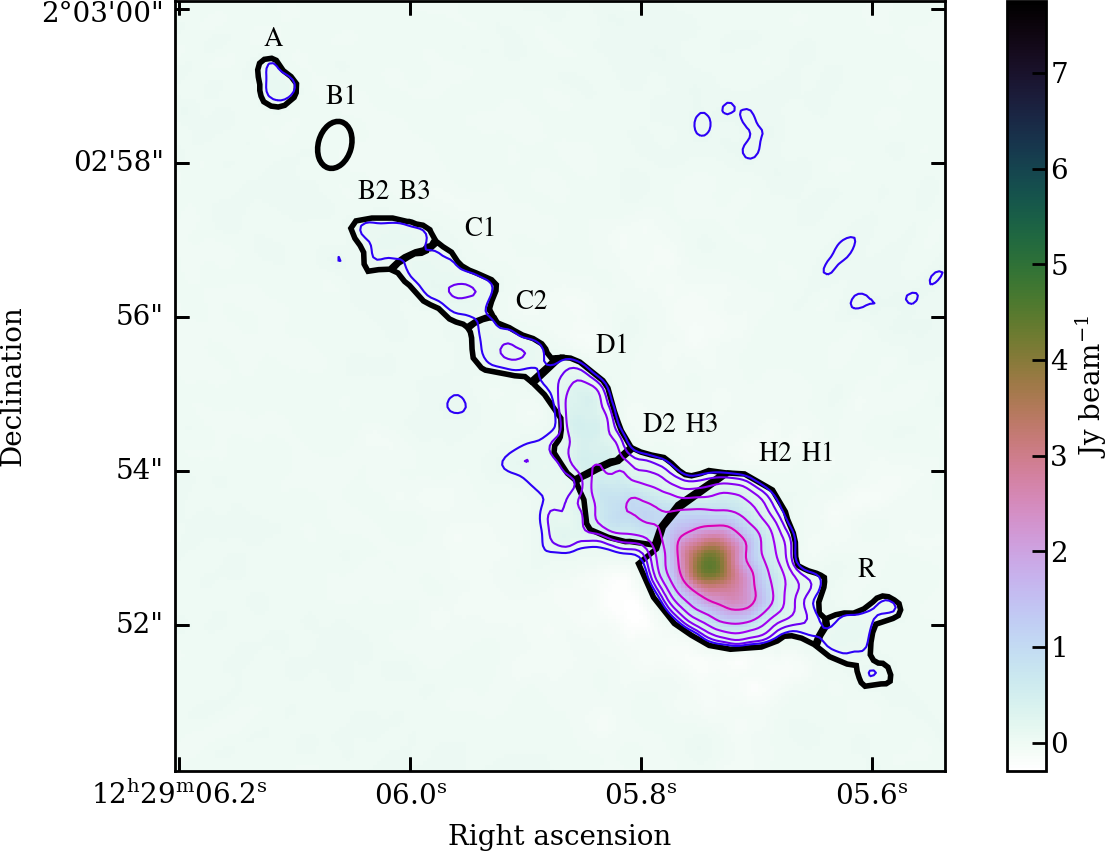}
    \caption{LOFAR images of 3C~273 at 151 MHz with a beam size of $0.31 \times\, 0.21$ arcsec. Contours levels are set such that $S_{cont} = 40\times2^{n}$ mJy beam$^{-1}$. Shown are the full source (left), the outer jet with the components labelled (right).}
    \label{fig:3C273-lofar-images-highres}
\end{figure*}

Due to the relatively high surface brightness of 3C~273, maximising image resolution rather than image sensitivity was the primary driver in determining the imaging parameters. Uniform weighting (robust $-2$) was therefore chosen to produce final images at the highest resolution, suitable for determining the properties of the small scale structure (e.g. knots). This provided a beam size of $0.31\times0.21$ arcsec with a position angle of $164 \degree$. The resulting radio maps are showing in Fig.~\ref{fig:3C273-lofar-images-highres}.

\subsection{Ancillary data} \label{sec:ancillary-data}

Higher frequency radio and near-infrared data were used to aid the LOFAR analysis and interpretation. The fully-calibrated VLA observations of \citet{2017A&A...601A..35P}
 were retrieved which include images from $0.3 - 43$ GHz at a range of resolutions. Near-infrared data taken with the Hubble widefield camera 3 (WFC3) were also obtained from the Hubble Legacy Archive\footnote{\url{https://hla.stsci.edu}} \citep{2006ASPC..351..406J} and are presented in Fig.~\ref{fig:3C273-hst}.

\section{Results and discussion}
\label{results}

\subsection{Morphology and brightness distribution}
\label{sec:morphology-of-the-jet}

The overall large-scale morphology of 3C~273 shown in Fig.~\ref{fig:3C273-lofar-images-highres} is in good agreement with previous studies performed with the VLA (e.g. \citealp{2017A&A...601A..35P}). While this suggests an overall good level of image fidelity has been achieved some unusual features on smaller scales are present in the vicinity of the core region, particularly in the north/south direction. Due to 3C~273's low declination, the uv tracks for 3C~273 are almost perfectly horizontal hence any slowly-varying errors of individual baselines will result in artefacts in the north/south direction as is observed in Fig.~\ref{fig:3C273-lofar-images-highres}. These features are therefore excluded from our analysis.

Determining the RMS noise from a blank region of sky well away from the source, we find that $\sigma_{rms} = 4.8$ mJy beam$^{-1}$. The peak brightness is located in the core region with a peak brightness of $7.8$ Jy beam$^{-1}$ giving a dynamic range of $1614$. This is lower than that obtained by \citet{2012A&A...547A..56D} who achieved a dynamic range of ${\sim}5000$ for their (non-international baseline) LOFAR observations of M87. While the sources are relatively close on the sky (with both located in the Virgo constellation), the declination of M87 is $10 \degree$ higher than 3C~273, with the M87 study also using a larger integration time and bandwidth ($116 - 162$ MHz) than for the observations used in this paper. The dynamic range is therefore at the expected level given the increased resolution provided by the inclusion of the international baselines, that is also impacted by the super-station resulting in an increased noise level in the image due to the loss of the core-to-core station baselines. To estimate the theoretical noise, we account for the loss of core-to-core baselines, uniform weighting for imaging, the amount of data flagging, and the low declination of the source to arrive at a value of $\sim0.23$mJy beam$^{-1}$. However, this value must be taken with a grain of salt as there are several important factors which we have not taken into account which could bridge the gap to the final $\sigma_{rms}$ we measure. First, and most critical, is the fact that M87 is $\sim$10 degrees away with a flux density in excess of 10$^3$ Jy at 151 MHz. Although demixing was performed to remove the contributions of this off-axis source, the results are only as good as the calibration in the direction of this bright off-axis source, and we do not yet have a high resolution model of M87 at these frequencies. Even optimistically, 1 percent residuals would still be $>10\,$Jy. It is crucial to note that the core stations are combined into the super-station after demixing but before self-calibration; any effects from improper subtraction of M87 during the demixing process will be `baked in' to the super-station, and these effects cannot be removed by subsequent self calibration. Better demixing of M87 is not possible until we have a high-resolution model, and even then will be computationally challenging as the quality of demixing scales with the number of components in the model source. Second, the combination of core stations is expected to increase the theoretical noise slightly due to the combined station beam, which is not yet well mapped. This effect is at the level of a couple percent for a high declination source, but we have not assessed this yet for low declination sources. Finally, the elevation correction we implemented only accounts for the projection of the stations and does not include complications such as the increased ionospheric path length at low declination.

\citet{2019A&A...622A...1S} show the LOFAR flux calibration error for HBA observations to be $\approx 15$ percent. Due to the additional core flux scaling performed, the observations presented within this paper are likely to improve on this value but we opt to use the $15$ percent to provide a conservative estimate. We therefore find a total flux density for the source of $91 \pm 14$ Jy. This agrees well with the previous VLBI study at $151$ MHz by \citet{2010A&A...520A..62A} who find a total flux density of $97.95 \pm 4.90$ Jy. We are therefore confident that our flux calibration is robust.

\begin{figure*}
    \centering
    \includegraphics[width=\textwidth]{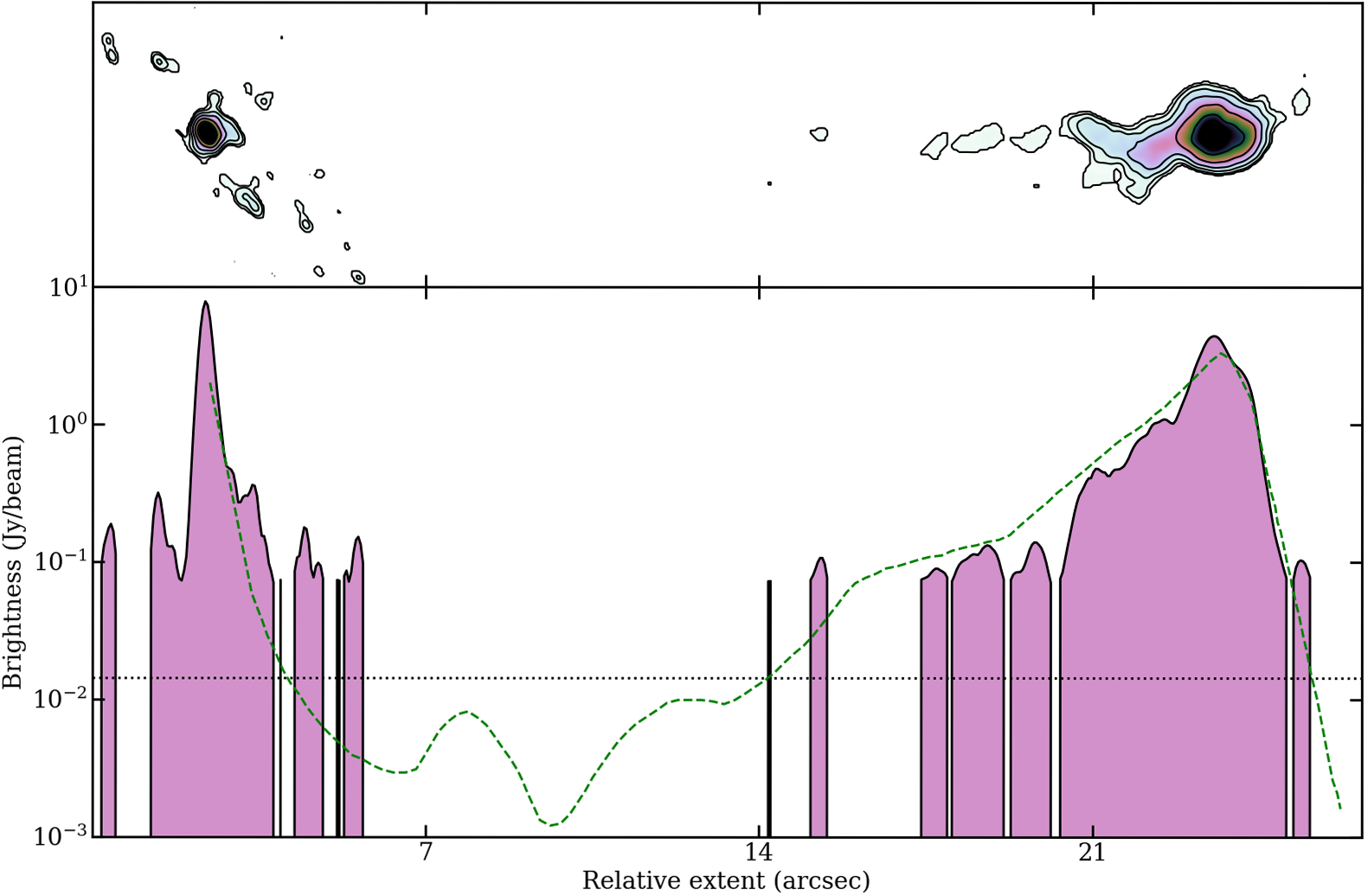}
    \caption{3C~273 ridgeline histogram with the brightness measured relative to the core. The image has been rotated counterclockwise by $42\degree$ such that the top of the image points northwest. The features around the core are likely to be imaging artefacts. The green dashed line shows the results of \citet{1985Natur.318..343D} and the dotted black line the 3$sigma$ RMS noise of the LOFAR image.}
    \label{fig:3C273-ridgeline-histogram}
\end{figure*}

The core has the highest peak brightness in the image ($7.8$ Jy beam$^{-1}$), which is $1.8$ times that of the next brightest feature, the radio hotspot where the jet terminates ($4.4$ Jy beam$^{-1}$). The relative brightness of the features can be seen in Fig.~\ref{fig:3C273-ridgeline-histogram}. While the inner jet is not visible, structure in the outer jet is apparent with all but one of the jet structures observed at optical wavelengths (Fig.~\ref{fig:3C273-knot-names-border}) identified in the LOFAR image (Fig.~\ref{fig:3C273-lofar-images-highres}, bottom). Knot B2 is the exception, for which we set an upper limit on its brightness of $40$ mJy beam$^{-1}$. A summary of the measured flux densities for each of the knots is shown in Table~\ref{tab:knot-fluxes}.

\begin{figure}
    \centering
    \includegraphics[width=\columnwidth]{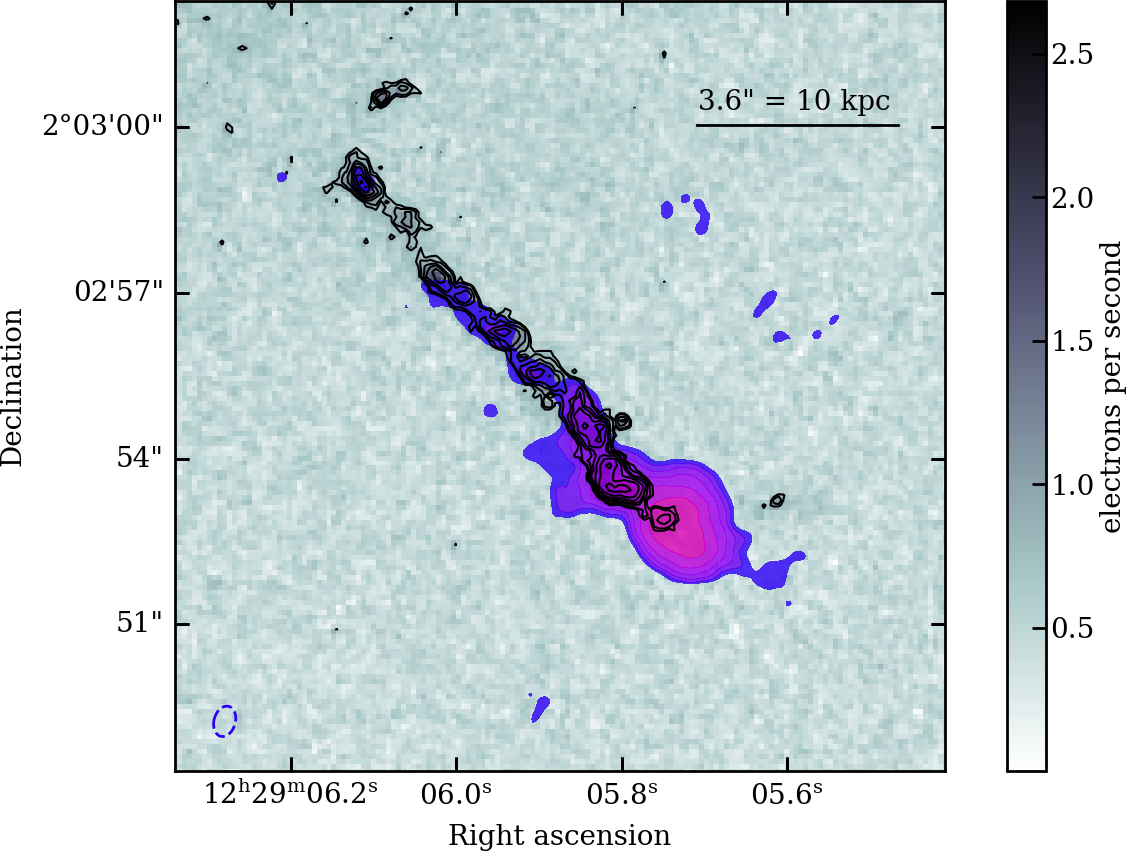}
    \caption{LOFAR data overlaid on HST near-infrared data. Black contours mark the HST emission and are set such that $S_{cont} = 0.7\times2^{n}$ e$^{-}$ s$^{-1}$. The filled contours show the LOFAR data set such that $S_{cont} = 40\times2^{n}$ mJy beam$^{-1}$.}
    \label{fig:3C273-contours-on-hst}
\end{figure}

Fig.~\ref{fig:3C273-contours-on-hst} shows the LOFAR images of the jet overlaid on HST data. The radio jet extends ${\sim} 2$ arcsec beyond the termination point of the jet with the shape of the radio jet closely following that of the optical jet. It is interesting to note that the galaxies In1 and In2 (Fig.~\ref{fig:3C273-knot-names-border}), while detected with HST, are not present in the LOFAR observation suggesting that these sources are radio quiet AGN.

\begin{figure*}
    \centering
    \includegraphics[height=6.8cm]{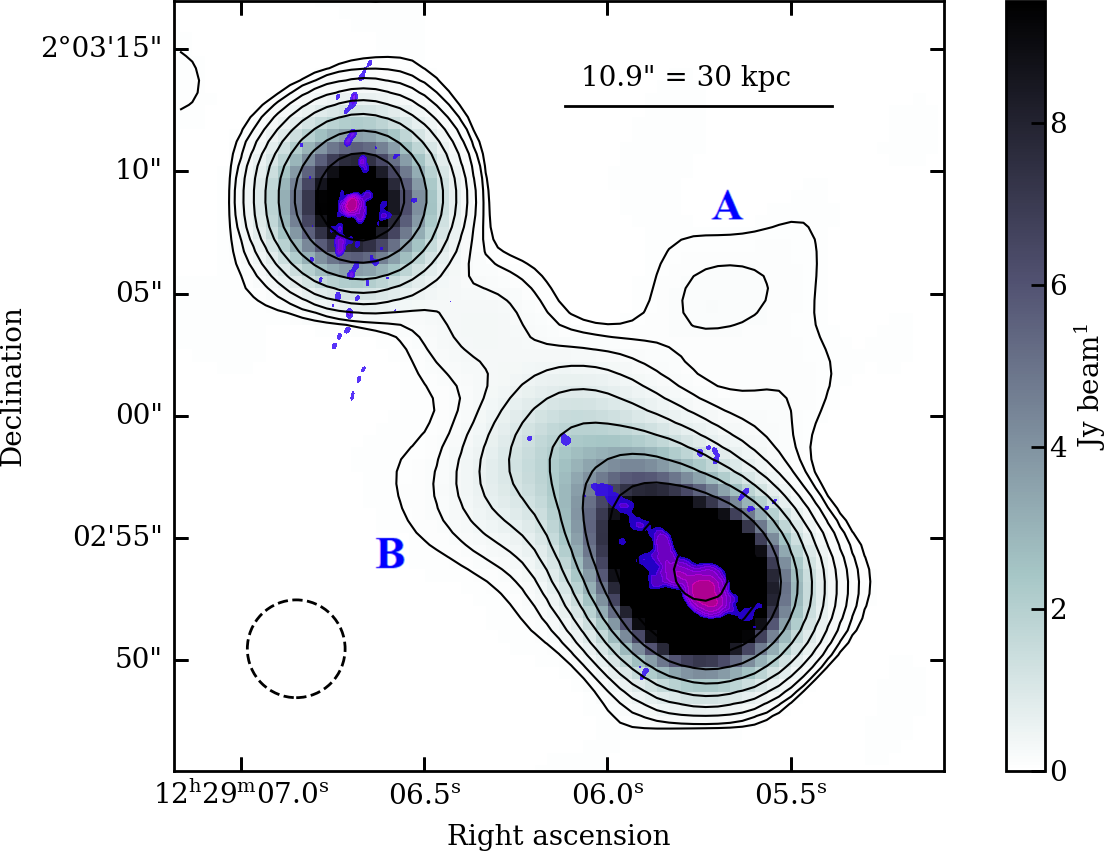}
    \includegraphics[width=\columnwidth]{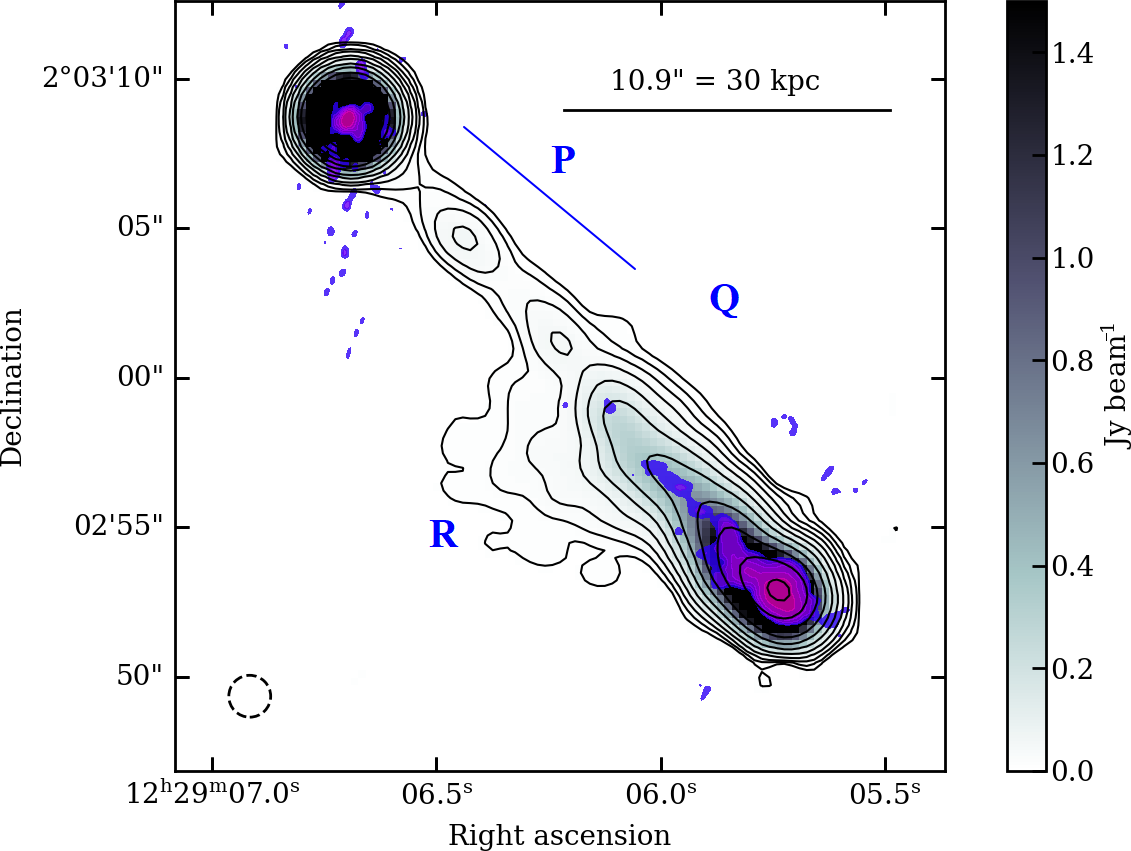}\\
    \vspace{2mm}
    \includegraphics[width=\columnwidth]{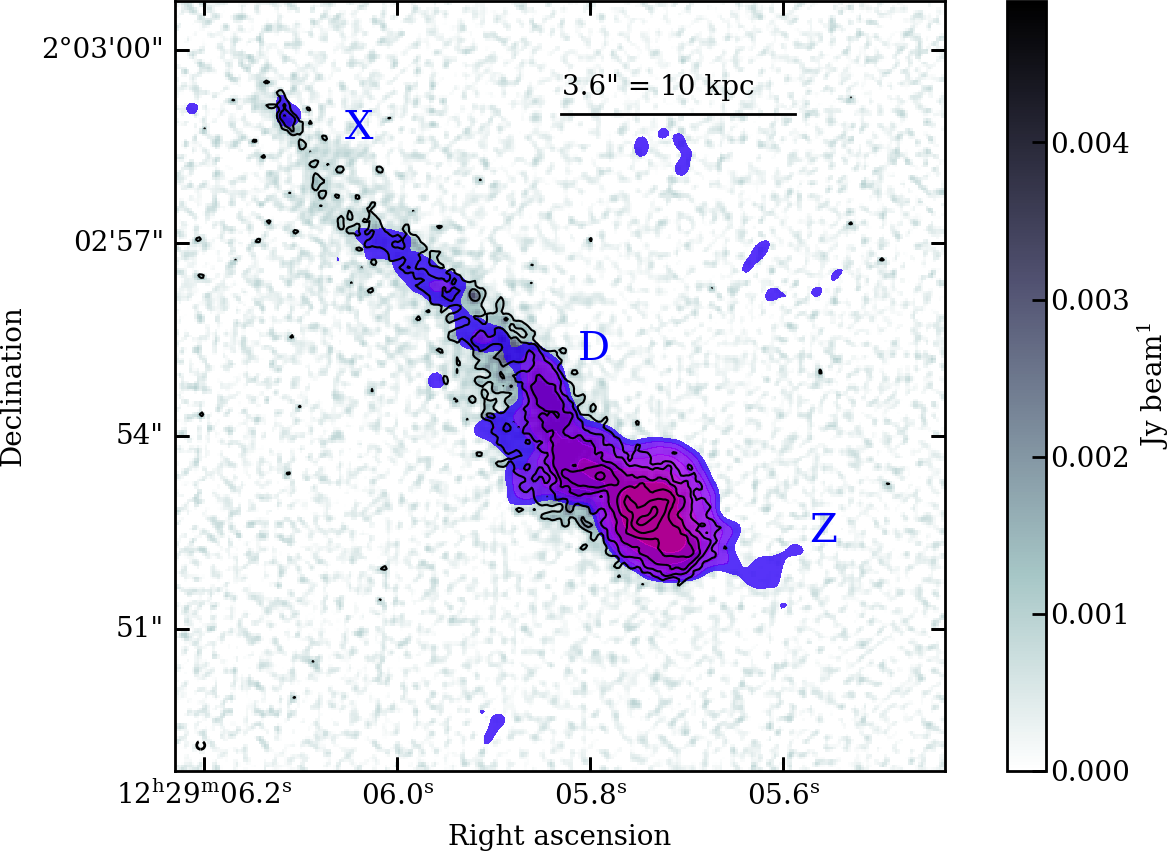}
    \caption{LOFAR data overlaid on VLA data taken from \citet{2017A&A...601A..35P} at P-, L-, and U-band (top left, top right, and bottom respectively). The beams of the VLA images are 4 arcsec (P-band), 1.4 (L-band), and 0.125 (U-band) arcsec and are shown by dotted circles. Black contours mark the VLA emission and are set such that $S_{cont} = \sigma_{rms}\times2^{n}$. The filled contours show the LOFAR data set such that $S_{cont} = 40\times2^{n}$ mJy beam$^{-1}$.}
    \label{fig:3C273-contours-on-vla}
\end{figure*}

Fig.~\ref{fig:3C273-contours-on-vla} shows the LOFAR images overlaid on VLA images at $325$ MHz, $1.4$ GHz, and $15$ GHz. The $325$ MHz P-band image is at $4$ arcsec resolution and illustrates the resolution that has been attainable prior to LOFAR at similar frequencies. At the VLA resolution, low-frequency radio lobes are detectable either side of the jet close to where the inner and outer jets meet (A and B on Fig.~\ref{fig:3C273-contours-on-vla}), and the lobe on the northern side of the jet (A) is particularly extended. For the VLA $1.4$ GHz L-band images at $1.4$ arcsec resolution, the inner jet is apparent in the VLA data (denoted P) and, while the radio lobe on the northern side of the jet is not detected (denoted Q), extended emission on the other side of the jet is present (denoted R). These inner jet features are not detected in the LOFAR images; however at VLA U-band frequencies ($15$ GHz) which provides a comparable resolution, the morphology of the jet is similar to that observed in the LOFAR data with knot B1 (denoted X) undetected in either image.

In the LOFAR data, we observe a minor extension to the south-west of the hotspot (denoted Z) with a peak brightness of $102$ mJy beam$^{-1}$ that is not detected in the VLA observation. The resolution between LOFAR and VLA at U-band frequencies is similar, but there are two orders of magnitude between the image frequencies. Measuring the RMS noise of the VLA image to be $0.3$ mJy beam$^{-1}$, the emission would require a spectral index steeper than $1.2$ to be detectable with LOFAR but not in the VLA U-band image. While from our observations it is not possible to deproject 3C~273, the observed one-sided jet and a hotspot-like jet termination suggest a \citet{1974MNRAS.167P..31F} class II (FR II) as a plausible morphology. In such cases, it has been shown that the injection index, the initial electron energy distribution at the point of acceleration, is commonly observed in the spectrum of FR-IIs to range between $\alpha_{inj} \approx 0.6$ and $1.0$ \citep{harwood13, harwood15, harwood16, harwood17a}. Given that at least some spectral ageing (the preferential cooling of high energy electrons resulting in spectral curvature) can be assumed, and the wide frequency range considered, a spectral index $>1.2$ is certainly achievable. While spectral steepening will be less significant at frequencies below $15$ GHz, causing this structure to be brighter in the P-band and L-band images, the resolution of the archival data is insufficient to discern structures on these scales. While it is therefore plausible that the extension is real, further investigation is needed to ascertain whether it is intrinsic to 3C~273, a background galaxy, or an imaging artefact.

\subsection{Spectral energy distributions of the knots} \label{sec:spectral-energy-distributions-of-the-knots}

\begin{figure*}
    \centering
    \includegraphics[width=\textwidth]{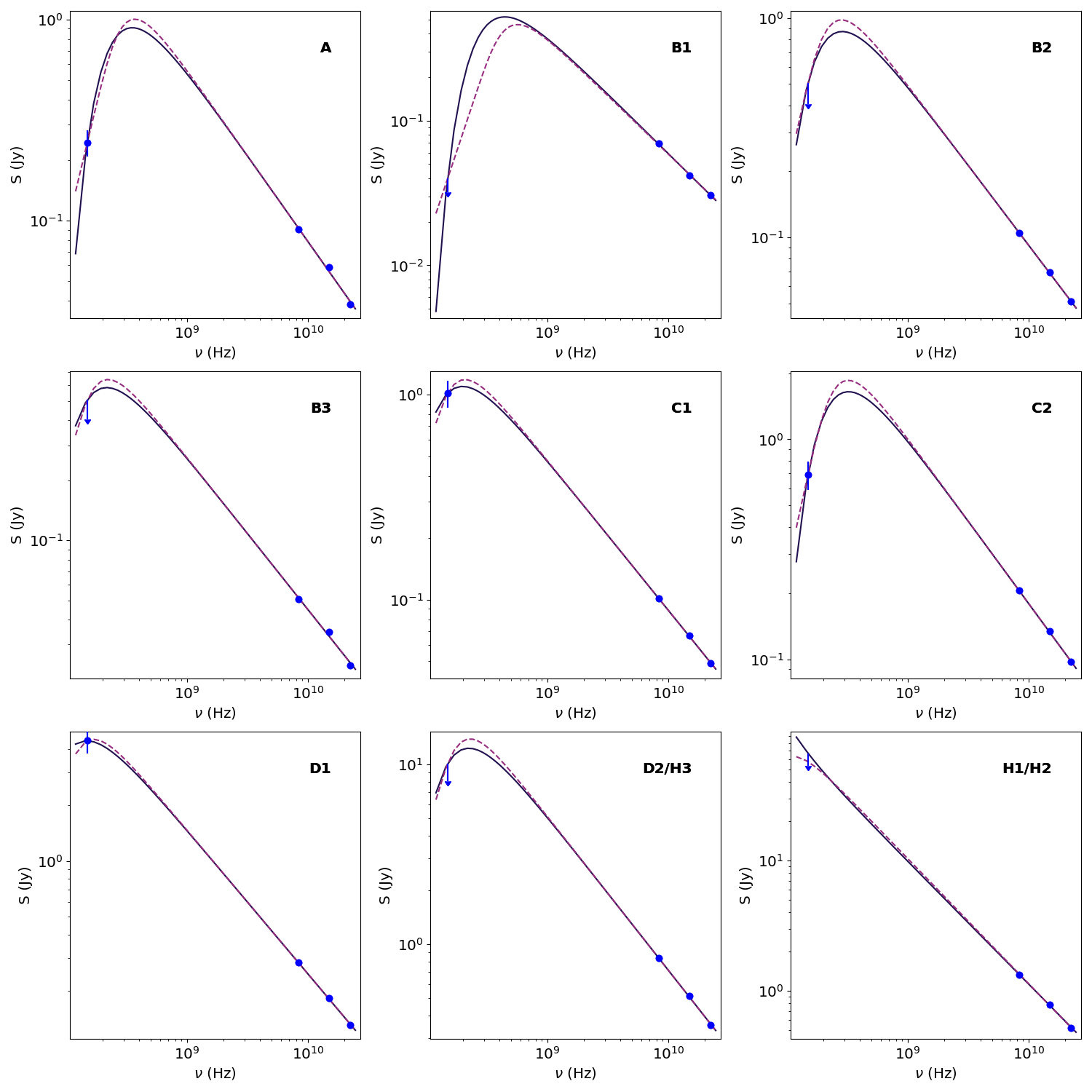}
    \caption{Spectral energy distributions of the knots in 3C273 with absorption models overlaid. The solid dark blue line indicates the FFA model and the dashed magenta line the SSA model. GHz data points taken from \citet{2007MNRAS.380..828J}.}
    \label{fig:jester07-fluxes}
\end{figure*}

The resolution of our observations provides the first opportunity to investigate the spectrum of the jet knots at low frequencies. Regions coincident with the optical features observed in Fig.~\ref{fig:3C273-knot-names-border} were defined for the radio images (Fig.~\ref{fig:3C273-lofar-images-highres}) and, combined with the VLA measurements of \citet{2007MNRAS.380..828J}, the spectral index for each region was calculated between $150$ MHz and $8.3$ GHz (Table~\ref{tab:knot-fluxes}). While the outer most regions (H1/H2) are inline with the expected spectrum of the hotspot and surrounding emission at low frequencies \citep{harwood16}, the origin of the spectrum of regions closer to the core are less clear. Three of the regions (A, B1, and C2) fall below the $\alpha=0.5$ physical minimum for first-order Fermi acceleration which is currently the favoured mechanism by which particles responsible for the radio emission are primarily accelerated. Two possible causes for the observed emission are free-free absorption (FFA) and synchrotron-self absorption (SSA) what can lead to a turnover in the spectrum at low frequencies. In both cases, attenuation of the emission is a function of frequency either through thermal absorption as a result of passing through an ionised screen (FFA), or due to the fact that brightness temperature cannot exceed the temperature of the non-thermal electrons (SSA). These models are discussed in detail in the context of radio-loud AGN elsewhere (e.g. \citealp{2015ApJ...809..168C, mckean16}) and so we do not repeat that process here, but for FFA the expected emission is given by (e.g. \citealp{kassim89})

\begin{equation}
    S_{FFA,\,\nu} = S_{\nu_0} \left(\frac{\nu}{\nu_0}\right)^{-\alpha} \exp{\left(-\tau_{\nu_{0}}(\nu/\nu_0)^{-2.1}\right)}
\end{equation}

\noindent where $S_{\nu_0}$ and $\tau_{\nu_{0}}$ is the flux density and optical depth at a reference frequency of $150$ MHz and $\alpha$ is the optically thin spectral index. Similarly, assuming the emission region is homogeneous, expected emission for the SSA model is given by \citep{kellermann66, tingay03}

\begin{equation}
    S_{SSA,\,\nu} = a \left(\frac{\nu}{\nu_{p}}\right)^{-\alpha} \left(\frac{1- e^{-\tau}}{\tau}\right)
\end{equation}

\noindent where

\begin{equation}
    \tau = \left(\frac{\nu}{\nu_{p}}\right)^{(\delta-4)/2}
\end{equation}

\noindent where $\nu_{p}$ represents the frequency at which the region becomes optically thick and $\delta$ is the power law index of the initial electron energy distribution such that $\alpha = (\delta-1)/2$. As measurements for the knots D2/H3 are combined by \citet{2007MNRAS.380..828J}, as are and B2/B3 in this paper, we are only able to provide upper limits for these regions in addition to knot B1 which is below our detection limit. Regardless of these limitations, we find that fitting of the FFA and SSA models are able provide a consistent interpretation of the observed spectra, the results of which are shown in Fig.~\ref{fig:jester07-fluxes} and Table~\ref{tab:ffassa}.

\begin{table}
    \centering
    \caption{Flux densities and spectral index values}
    \label{tab:knot-fluxes}
    \begin{tabular*}{\linewidth}{c@{\extracolsep{\fill}}ccc}%
        \hline
        \hline
        Feature & Flux density (Jy) & $\alpha_{150}^{8330}$ \\
        \hline
        A                & $0.245 \pm 0.037$          & $0.25 \pm 0.04$           \\
        B1               & $<0.040$                   & ---                        \\
        B2/B3            & $0.507 \pm 0.076$          & $0.29 \pm 0.04$           \\
        C1               & $1.02 \pm 0.152$          & $0.57 \pm 0.09$           \\
        C2               & $0.691 \pm 0.104$          & $0.30 \pm 0.05$           \\
        D1               & $4.44 \pm 0.67$          & $0.69 \pm 0.10$           \\
        D2/H3            & $10.09 \pm 1.51$         & $0.62 \pm 0.09$           \\
        H1/H2            & $66.80 \pm 10.02$        & $0.97 \pm 0.15$           \\ 
        \hline
    \end{tabular*}
    \vskip 5pt
    \begin{minipage}{\columnwidth}
        \small Flux densities at $150$ MHz and the $150 - 8330$ MHz spectral index values for the jet structures shown in Fig.~\ref{fig:3C273-lofar-images-highres}.
    \end{minipage}
\end{table}

The H1/H2 region is inline with expectations for being the primary acceleration region (i.e. the hotspot), being well described within errors by a power law and is optically thin at $150$ MHz in both the FFA ($\tau_{{\nu}_{0}} = -0.13 \pm 0.20$) and SSA ($\nu_{p} = 97.86 \pm 2.47$ MHz) models. The optically thin spectral index component is relatively consistent for both models and across all knots, being dominated by the GHz observations which are well above the frequency at which the regions becomes optically thick. For regions closer to the core where the observed brightness is significantly lower than would be expected at $150$ MHz, the models are able to provide a good description of the observed spectrum in all cases. For knot A, where we are not constrained by upper limits, we find our LOFAR observations are likely just below the frequency at which the region becomes optically thick ($\nu_{p} = 311.3 \pm 17.2$ MHz) with an optical depth of $\tau_{{\nu}_{0}} = 2.45 \pm 0.20$.

The wide variation in $\tau_{{\nu}_{0}}$ across the knot regions suggest that, if FFA is the mechanism by which absorption is occurring, the ionised screen cannot be homogeneous. While an inhomogeneous screen external to 3C~273 is possible, the scales on which the jet and knots are observed suggests the absorption is likely occurring in situ. Such a local ionised medium is perhaps not surprising given the prominent nature of the observed knots. The jets of FR-IIs typically remain uninterrupted out to large distances and interaction between, for example, the jet sheath and an inhomogeneous ionised medium may explain the prominent knotty structure of the jet at radio frequencies. 

With only one data point at low frequencies it is currently not possible to differentiate between the two models; however both the FFA and SSA models are able to provide a good description for all the knots. We therefore suggest that, while the precise mechanism is not clear, absorption is present within the knot regions and should be carefully considered when undertaking future investigations. The planned upgrade of the LOFAR Low Band Antennas (LBA) will provide observations at ${\sim50}$ MHz that will better constrain these models and may also be able to differentiate between the two models for the most heavily absorbed knots. Ultimately, Square Kilometre Array (SKA) -- GMRT VLBI will provide the observations capable of constraining the spectrum at a few hundred MHz where the difference between SSA and FFA is most significant. Upcoming and future instruments, when combined with LOFAR results, will be key to continuing this form of investigation.

\begin{table}
    \centering
    \caption{Free-free and synchrotron self absorption model fitting values}
    \label{tab:ffassa}
    \begin{tabular*}{\linewidth}{c@{\extracolsep{\fill}}ccccc}%
        \hline
        \hline
        Feature & $\tau_{\nu_{0}}$& $\alpha_{FFA}$ & $\nu_{p}$ & $\alpha_{SSA}$\\
        &&($\pm 0.03$)&(MHz)&($\pm 0.03$)\\
        \hline
        A                & $2.45 \pm 0.20$      & $0.85$       & $311.3 \pm 17.2$       & $0.85$           \\
        B1               & $>3.56$              & $0.82$       & $>470.1$               & $0.81$           \\
        B2               & $>1.360$             & $0.73$       & $>226.9$               & $0.73$           \\
        B3               & $>0.78$              & $0.76$       & $>180.6$               & $0.76$           \\
        C1               & $0.63 \pm 0.20$      & $0.73$       & $167.2 \pm 19.2$       & $0.73$           \\
        C2               & $1.80 \pm 0.20$      & $0.75$       & $260.9 \pm 15.5$       & $0.75$           \\
        D1               & $0.36 \pm 0.20$      & $0.77$       & $138.6 \pm 19.6$       & $0.77$           \\
        D2/H3            & $>0.95$              & $0.85$       & $>192.8$               & $0.85$           \\
        H1/H2            & $-0.13 \pm 0.20$     & $0.94$       & $97.86 \pm 2.47$       & $0.96$           \\
        \hline
    \end{tabular*}
    \vskip 5pt
    \begin{minipage}{\columnwidth}
        \small FFA and SSA model fitting results for the jet structures shown in Fig.~\ref{fig:3C273-lofar-images-highres}. $\tau_{\nu_{0}}$ is the FFA optical depth at a reference reference of 150 MHz, $\nu_{p}$ the frequency at which the region becomes optically think for the SSA model, and $\alpha$ the optically thin spectral index.
    \end{minipage}
\end{table}

\subsection{Estimate of the diffuse emission} \label{sec:estimate-of-the-diffuse-emission}

To determine the flux density of the diffuse emission surrounding the knots, we blanked the entire image at 150 MHz below a threshold of $5\sigma_{rms}$ and summed over the pixels containing the core and jet emission (thus avoiding bright artefacts). Subtracting this from the total flux density of 91 Jy, we arrive at an estimate of $2.6 \pm 0.4$ Jy for the diffuse flux density.

To estimate the diffuse emission, the flux density of the entire jet was measured to a threshold of $10\sigma_{rms}$ and the sum of the knots flux densities subtracted, providing an estimate for the diffuse flux density of $2.8 \pm 0.4$ Jy. Some caution should be exercised with respect to the robustness of this value as it is possible that the diffuse emission is overestimated due to non-Gaussian shoulders present in the PSF. Conversely, due to observing the jet at a small angle to the line-of-sight, a significant fraction of the diffuse emission is likely to be in superposition with knots. This will act to both reduce the measured diffuse emission and increase the knot emission. Due to the relative brightness of the compact knots compared to the surrounding diffuse emission, this is unlikely to have a significant impact on the knot values discussed in Section~\ref{sec:morphology-of-the-jet} but will here act to reduce the amount of intrinsic diffuse emission we are able to measure. Determining the relative impact of each of these factors would require detailed modelling of the source and so is beyond the scope of this paper; however, we can make comparison to previous studies where such an estimate has been made.

\citet{2016ApJ...833...57P} measured the diffuse flux density at $327$ MHz to be $1.33 \pm 0.13$ Jy and extrapolated this to $151$ MHz to arrive at an estimate of $2.7$ Jy. Their image at $327$ MHz had a resolution of $7 \times 6$ arcsec so the core, jet, and diffuse emission were not well-resolved. \citet{2016ApJ...833...57P} determined that the diffuse flux density was underestimated by $50 - 100$ per cent and increased the value accordingly. In the LOFAR data, however, the core, jet, and diffuse emission are well-resolved so we make no adjustments to the calculated flux density. Nevertheless, the range of jet powers still overlaps with the final estimates of \citet{2016ApJ...833...57P} and so the more naive methods used in both papers appear to provide a reasonable estimate of the observed emission.

To convert optically-thin $150$ MHz flux densities from the lobes into estimates of the long term time-averaged jet power, $Q$, we apply the relations of  \citet{1999MNRAS.309.1017W} and \citet{2005ApJ...623L...9P} who find
\begin{equation}
    Q \approx \left(\frac{f}{15}\right)^{1.5} (1.1 \times 10^{45}) \, (X^{1+\alpha} \, Z^2 \, S_{150})^{0.857} \label{eq:q-upper}
\end{equation}
\vspace{-5mm}
\begin{multline}
    Z \equiv 3.31 - 3.65 \, (X^4 - 0.203 X^3 + 0.749 X^2 + 0.444 X + 0.205)^{-0.125} \nonumber
\end{multline}
\begin{equation}
    X \equiv 1 + z \nonumber
\end{equation}

\noindent where $z$ is the redshift and $f$ is an empirical multiplicative factor incorporating uncertainties associated with departures from minimum energy and variations in geometric effects, filling factors, hadronic contributions, and the low-frequency cutoff. \citet{2016ApJ...833...57P} state that $f$ is likely to be $10$--$20$ and so we therefore set $f = 20$, treating this as an upper limit. \citet{2005ApJ...623L...9P} independently derive an estimator for the jet power, where the lobe energy is primarily inertial (i.e. thermal, turbulent, and kinetic energy). This assumption was validated by X-ray data that indicate the energy in radio lobes is dominated by inertial energy, as opposed to magnetic field energy, unlike the hotspots that are usually near equipartition. \citet{2005ApJ...623L...9P} show that
\begin{gather} 
    Q \approx (5.7 \times 10^{44}) \times (X^{1+\alpha} \, Z^2 \, S_{150}) \label{eq:q-lower}
\end{gather}
\noindent represents a lower estimate on the jet power. Following the methodology of \citet{2016ApJ...833...57P}, we find that between the limits given by Eq.~\ref{eq:q-upper} and \ref{eq:q-lower} where $f = 20$, the jet power of 3C~273 is in the range $3.5 \times 10^{43}$ - $1.5 \times 10^{44}$ erg s$^{-1}$. This is in agreement with \citet{2016ApJ...833...57P}, who derive the jet power of $0.7 - 3.7 \times 10^{44}$ erg s$^{-1}$ which falls within the typical range for radio-loud quasars.

\subsection{Presence of the counterjet} \label{sec:presence-of-the-counterjet}

Despite the increased sensitivity and excellent uv coverage including short baselines provided by LOFAR observations the counterjet remains undetected. Assuming intrinsically symmetrical jets, where the observed asymmetry is due to relativistic beaming, from the $3\sigma_{rms}$ local to the expected counterjet region we determine an upper limit of $S_\mathrm{cj} = 40$ mJy beam$^{-1}$.

Using the counterjet peak brightness limit along with the measured peak brightness of the jet discussed in Section~\ref{sec:morphology-of-the-jet} we are also able to provide constraints on the bulk speed and the angle to our line-of-sight. For a a pair of relativistic jets, where one jet is moving towards the observer ($j$) and one is moving away ($cj$), the ratio of the observed flux densities is given by
\begin{equation}
    \frac{S_\mathrm{j}}{S_\mathrm{cj}} = \left( \frac{1 + \beta \cos \theta}{1 - \beta \cos \theta} \right)^{m - \alpha}\label{eq:oheeacj}
\end{equation}
\noindent where $\beta \in [0, \, 1]$, $\theta \in [0, \, \pi / 2]$, $m = 3$ for a knotty jet, and $\alpha = 0.8$ for optically-thin synchrotron emission. Substituting for our values of $4400$ and $40$ mJy beam$^{-1}$ for the jet and counterjet respectively, we find for 3C~273 that $\beta \cos \theta \geq 0.55$.

Assuming an inclination angle of $5.5\degree$, the midpoint of range determined by \citeauthor{2016ApJ...818..195M} (\citeyear{2016ApJ...818..195M}; $3.8 - 7.2 \degree$), we are able to place a lower limit on the bulk speed of $\beta \gtrsim 0.55$. This lower limit is in line with studies by \citet{2016ApJ...818..195M, 2017Galax...5....8M}, where it was asserted that the bulk motion of the kiloparsec-scale jet is mildly relativistic at most. Using the standard equation for the bulk Lorentz factor
\begin{equation}
    \Gamma = \frac{1}{\sqrt{1 - \beta ^ 2}} \, . \label{eq:gammaloretnz}
\end{equation}
for $\beta \geq 0.55$ for 3C~273 we find that $\Gamma \geq 1.2$. As discussed in Section \ref{intro}, The IC/CMB mode requires significant bulk velocities on kiloparsec scales ($\Gamma \approx 10$) which cannot be excluded by this lower limit, although a smaller angle to the line-of-sight is likely still required if such an emission mechanism were to dominate.

While we are able to provide some constraints on the jet properties using the upper limits given above, the absence of any counterjet emission raises some interesting questions. Given the steep spectrum of lobe and hotspot emission, LOFAR is arguably the instrument best placed to make such detection given its UV coverage and resolving power at low-frequencies. The brightest regions of the jet (the H1/H2 lobe) should have an analogous counterpart if jet symmetry is assumed that should be the least impacted by beaming effects in the counterjet from the observers frame of reference. It may be the case that the counterjet emission remains below the detection limit of our observations although with such a consistent lack of detection across all wavelength, alternative interpretations should be considered.

One possible cause is that while the jets may be intrinsically similar, light travel effects may cause an observational asymmetry. Assuming the inclinations determined by \citep{2016ApJ...818..195M} described above, the deprojected distance between the furthest extent of the two jets is between $479$ and $905$ kpc, which translates to between a $\approx 1.6$ and $3.0$ Myr light travel time. It is therefore possible that the counterjet is not observationally symmetric due to being observed at an earlier time and the location and/or morphology of the peak emission may differ from that of the main jet.

Another possibility that cannot be ruled out is that the counterjet is located behind the core, relative to our line-of-sight. This may be due to either a slight asymmetry in the jet launch angle or bending of the jet on kiloparsec scales. The narrow angle to the line-of-sight means that only a relatively small deviation would be required to cause the brightest outer lobe regions to be obscured by the core and so provides a plausible explanation for the non-detection of the counterjet.

As is often the case, the true cause is likely some combination of beaming and light travel time effects, with possibly some jet asymmetry that is the cause of the persistent non-detection of the counterjet. It is beyond the scope of this paper to determine which, if any, of these scenarios is the dominant cause, but the lack of detection at the low-frequencies suggests interpretations other than a simple lack of sensitivity in the observations should be considered. Deeper observations using the the LOFAR HBA ($150$ MHz) and at lower frequencies using the LOFAR LBA (${\sim50}$ MHz) would prove interesting in this regard by either detecting, or further constraining, the counter-jet's peak brightness.

\section{Conclusions}
\label{conclusions}

In this paper we have presented the first high-resolution observations of 3C~273 at MHz frequencies, recovering both compact and diffuse emission on sub-arcsecond scales. We have shown that robust, high-fidelity imaging of low-declination complex sources is possible with the LOFAR international baselines and that such observations are key to understanding the physics that underlie the small-scale structures of radio-loud AGN. By analysing the kpc scale morphology and spectrum of 3C~273 at MHz frequencies to answer the questions posed in Section \ref{intro} we find that:

\begin{enumerate}
\item The small-scale structure of 3C~273 matches that at higher frequencies.\\
\item The low-frequency spectrum of the jet knots can be well described by free-free absorption (FFA) and synchrotron self-absorption (SSA) models.\\
\item We derive a kinetic power for the jet in the range $3.5 \times 10^{43}$ - $1.5 \times 10^{44}$ erg s$^{-1}$ which agrees with previous estimates derived from higher frequency observations.\\
\item The counterjet remains undetected at $150$ MHz, placing limits on the peak brightness of $S_\mathrm{cj\_150} < 40$ mJy beam$^{-1}$.\\
\item We derive lower limits of $\beta \gtrsim 0.55$ and $\Gamma \geq 1.2$ for the for the bulk speed and  Lorentz factor respectively.
\end{enumerate}

Further low-frequency investigations (e.g. using LOFAR LBA observations) are required to determine many of the details of 3C~273, such as counterjet detection and to better differentiate between absorption models. However, our investigation has made the first steps towards understanding the detailed mechanics of 3C~273 at these wavelength and provided the first constraints on the characteristics of 3C~273 at low-frequencies; a step that will be vital moving forward with both the LOFAR international baselines and towards the SKA era.

\section*{Acknowledgements}
\label{acknowledgements}
We wish to thank the referee, Rick Perley, for his constructive comments and suggestions which have helped improve this paper. SM acknowledges support from the Irish Research Council’s Government of Ireland Postgraduate Scholarship scheme. LKM is grateful for support from the UKRI Future Leaders Fellowship (grant MR/T042842/1). CG acknowledges support from the ERC Starting Grant ClusterWeb 804208. JM acknowledges financial support from the State Agency for Research of the Spanish MCIU through the ``Center of Excellence Severo Ochoa'' award to the Instituto de Astrof\'isica de Andaluc\'ia (SEV-2017-0709) and from the grant RTI2018-096228-B-C31 (MICIU/FEDER, EU).

\bibliographystyle{aa}
\bibliography{references}

\label{lastpage}
\end{document}